\documentclass[aip,reprint]{revtex4-1}
\usepackage{amsfonts,amsmath,amssymb,amsthm,float}

\newcommand{\RR}{{\mathbb R}}
\newcommand{\CC}{{\mathbb C}}

\newtheorem{defn}{Definition}[section]

\draft 

\begin{document}


\title{Derivation of Quantum Mechanics algebraic structure from invariance of the laws of Nature under system composition and Leibniz identity} 



\author{Florin Moldoveanu}
\affiliation{Logic and Philosophy of Science Research Group, University of Maryland at College Park}


\date{\today}

\begin{abstract}
Products and tensor products are linked by a universal property. Imposing the invariance of the laws of Nature under tensor composition along with Leibniz identity determines quantum and classical mechanics algebraic structure through the interplay between products, coproducts, and the tensor product. Violations of Bell's inequalities distinguishes quantum from classical mechanics. \end{abstract}


\maketitle 

\section{Introduction}

Recently there has been extensive interest in deriving quantum mechanics from physical principles \cite{BarnumWilce, DakicBrukner, MasanesQM, ChiribellaQM, HardyQM}. Common in most of those approaches are the usage of composition axioms:

\begin{itemize}
\item Barnum and Wilce \cite{BarnumWilce}: ``composites are locally tomographic'',
\item Dakic and Brukner \cite{DakicBrukner}: ``the state of a composite system is completely determined by local measurements on its subsystems and their correlations'',
\item Lluis Masanes and Markus Muller \cite{MasanesQM}: ``the state of a composite system is characterized by the statistics of measurements on the individual components'',
\item Chiribella, D'Ariano, and Perinotti \cite{ChiribellaQM}: ``if two states of a composite system are different, then we can distinguish between them from the statistics of local measurements on the component systems'',
\item Lucien Hardy \cite{HardyQM}: ``Composite systems rules" ($ N_{A\otimes B} = N_A N_B $ and $ K_{A\otimes B} = K_A K_B $ where $N$ is the dimension of the state space and $K$ are the number of degrees of freedom).
\end{itemize}

It is easy to understand why composition plays a role. Quantum mechanics achieves correlations higher than the Bell limit \cite{Bell1, AspectExperiment}. If we want to distinguish between classical and quantum mechanics, considerations of composite systems must be included in one form or another. However a 1976 result by Emile Grgin and Aage Petersen \cite{GP} hints that composition arguments are much more powerful because using composition they managed to constructively obtain an algebraic relationship obeyed by quantum mechanics. 

Composition arguments can achieve more because there is a universal property between products and tensor products which reduce bilinear algebraic products to linear operators on the tensor product \cite{UnivesalProperty}. As such any requirements we impose on composition constrain the products involved in the algebraic structure as well. 

Following Grgin and Petersen, we will consider the invariance of the laws of Nature under composition \cite{GP}. In addition to this we will demand the functoriality of time translation because the laws of Nature do not change under time evolution. From this in the infinitesimal case we will derive the Leibniz identity. We will show that the nontrivial interplay between products, coproducts, and the tensor product completely determines the algebraic structure of quantum and classical mechanics.

The algebraic approach to quantum mechanics was originally introduced due to the mathematical difficulties of quantum field theory, in particular the lack of a Hilbert space for certain problems. Citing Emch: ``The basic principle of the algebraic approach is to avoid starting with a specific Hilbert space scheme and rather to emphasize that the {\em primary objects} of the theory are the fields (or the observables) considered as purely algebraic quantities, together with their linear combinations, products, and limits in the appropriate topology.'' \cite{EmchBook}. The usual algebra considered is the C*-algebra of bounded operators from which the Gelfand-Naimark-Segal (GNS) construction \cite{GNSReference} recovers the Hilbert space formulation. However the full algebraic mathematical structure of the algebraic formalism is the Jordan-Lie algebra for quantum mechanics and the Poisson algebra for classical mechanics \cite{LandsmanBook} because a quantum mechanics C*-algebra can be further decomposed into the Jordan algebra of Hermitean operators (given by the symmetric product which preserves hermiticity), and the Lie algebra of anti-Hermitean operators (given by the commutator). The compatibility relationship between the two algebras allows the C*-algebra to be associative. 

Composition of two physical systems obeys the properties of a commutative monoid \cite{GP} and composition and information are naturally expressed in the category theory formalism. Category theory was successfully used in the past to introduce a pictorial representation of quantum information \cite{AbramskyCoecke}. 

\section{Strategy of the approach}

We will start from two physical principles: the invariance of the laws of Nature under composition and the functoriality of the time translation operator.  

The physical principle of the invariance of the laws of nature under composition can be interpreted in several ways. First, it means that if system $\mathcal{A}$ is described by a theory of nature, system $\mathcal{B}$ is described by the same theory of nature, then the composed system $\mathcal{A}\boxtimes \mathcal{B}$ is described again by the same theory of nature. Second, it shows that the dynamical laws do not change with increased degrees of freedom. Last, it can be interpreted as the impossibility to have within our universe a ``pocket universe'' with distinct laws of nature.
 
For the second physical principle, if $T$ is a time translation operator we demand $T$ to be a functor. This means that the laws of Nature do not change under time evolution. In particular time translation is preserving all algebraic products:

\begin{equation*}
T (A \circ B) = T(A) \circ T(B),
\end{equation*}

\noindent where $\circ$ stands for any bilinear product which may be involved in the description of our physical system (commutator, Jordan product, Poisson bracket, regular function multiplication, Moyal bracket, etc).

In order to generate mathematical consequences we consider infinitesimal time translation $T = I + \epsilon \alpha$ where $\alpha$ is the commutator in quantum mechanics or the Poisson bracket in classical mechanics. The reason for this restriction is that we need to start with a bilinear product $\alpha$ which can impose a linear operation in the tensor product.

In the infinitesimal limit, the time translation yields the Leibniz identity:

\begin{equation*}
A\alpha (B \circ C) = (A\alpha B) \circ C + B \circ (A \alpha C).
\end{equation*}

The Leibniz identity works in both commutative (phase space) settings as well as in non-commutative (Hilbert space) settings and we do not want to attach any ontological interpretation to it or the domain or codomain of the product $\alpha$ because category theory arguments are independent from the nature of the objects. From the usual formulations of quantum mechanics we know that the product $\alpha$ acts on either the operators in a Hilbert space (where $\alpha$ is the commutator) or on the functions in phase space (where $\alpha$ is the sine-Moyal bracket), but for our purposes we want to treat it formally as an abstract algebraic operation and postpone the discussion of concrete representations to the end after exhausting all category theory arguments first. As a finer point of discussion we can either justify the Leibniz identity from the functoriality of time translation, or we can simply postulate its existence.  

In the end we will obtain only three possible solutions corresponding to three classes of composition: elliptic (quantum mechanics), parabolic (classical mechanics), and hyperbolic (hyperbolic quantum mechanics \cite{KhrennikovSegre}). The hyperbolic case is unphysical because in this case one cannot create a state space able to generate physical predictions. Nature can only be in one of the two remaining composition classes, and by experimental evidence \cite{AspectExperiment} Nature is quantum mechanical.

\section{Composing two physical systems}
In their 1976 paper \cite{GP}, Grgin and Petersen introduced a so-called ``composition class'' $\mathcal{U}$ which describes the composition of two physical systems $\mathcal{S}_1$ and $\mathcal{S}_2$: $\mathcal{S}_1 \boxtimes \mathcal{S}_2$. For quantum mechanics $\boxtimes$ is the tensor product $\otimes$ of Hilbert spaces, while for classical mechanics $\boxtimes$ is the Cartesian product of phase spaces \cite{AKapustin}. In the algebraic approach system composition is uniformly expressed for both quantum and classical mechanics by the tensor product $\otimes$ and this shows the advantage of this approach over the state/phase space formalism for our purposes. $\mathcal{U}$ has the properties of a monoid: associativity and the existence of a unit. The justification of associativity is that whenever we can perform experiments, the separation of a physical system into subsystems and the reverse process of composition reflect the boundary between what is it measured and what it is not. The existence of a unit is a consequence of the relational character of the laws of Nature. For example in phase space the Poisson bracket is unaffected by constant functions and the dynamic is not affected by the ground energy level. We will use the composition class approach to completely determine the algebraic properties of quantum and classical mechanics.

\subsection{$k$-algebras and $k$-coalgebras}
From the bilinear product $\alpha$ respecting the Leibniz identity we can construct an algebra $A$ over a field $k$ (with $k=\RR$ for phase space formulation, and $k=\CC$ for Hilbert space formulation) which we will call a {\em k-algebra} \cite{SweedlerBook} $(A, \alpha, u)$ with $A$ a $k$-vector space, $\alpha$ the associative product, and $u$ the unit $u: k \rightarrow A$, subject to the usual associativity and unitary diagrams, and where $A\otimes A \xrightarrow{\alpha} A$ is a linear product. Also $k\otimes A \rightarrow A$ and $A\otimes k \rightarrow A$ are the natural isomorphisms. For convenience, by abuse of language, we will not distinguish between the original bilinear product $\alpha$ and the new linear product $\alpha$ because it will be clear from the context which product we will be talking about.

The reason for this formulation is twofold. First, by a universal property we convert the bilinear product into a linear product where we can use system composition arguments. Second, in this formulation the idea of a coproduct and coalgebra appears naturally. A $k$-coalgebra is the same as a $k$-algebra but with all arrows reversed. In this case the product is called a coproduct, and the unit is called a counit. The $k$-coalgebra formalism will help define the product $\alpha$ for a bipartite system.

\subsection{The necessity of a second product}
Let us start with a preliminary result which we will repeatedly use. We want to show that $(1\alpha 1) = 0$ when $\alpha$ obeys the Leibniz identity. The proof is as follows. We start with $1\alpha (1\alpha f)$ and use the Leibniz identity:

\begin{equation*}
1\alpha (1\alpha f) = (1\alpha 1) \alpha f + 1\alpha (1\alpha f),
\end{equation*}

\noindent from which we get $(1\alpha 1) \alpha f = 0$. Because this is valid for any $f$ implies in turn that $1\alpha 1 = 0$.

Now suppose that we have a toy theory (not describing Nature) in which only one algebraic product $\alpha$ is allowed to exist. The product $\alpha$ is defined in two physical systems $1$ and $2$ and we want to define the product $\alpha_{1\otimes 2}$ for the composed system $1\otimes 2$. The only way we can do this is as follows:

\begin{equation*}
(f_1\otimes f_2) \alpha_{1\otimes 2} (g_1\otimes g_2) = a (f_1\alpha g_1) \otimes (f_2\alpha g_2),
\end{equation*}

\noindent where $f_1, g_1 \in A$ for system $1$, $f_2, g_2 \in A$ for system $2$, $\alpha$ defined in $A$, and $a \in k$ is a constant parameter.  

Now we can exploit the natural isomorphism and the existence of a unit element for the composition class. If we pick $f_2 = g_2 = 1 \in k$, we have:

\begin{equation*}
(f_1 \otimes 1) \alpha_{1\otimes 2} (g_1 \otimes 1) = (f_1 \alpha g_1) = (f_1 \alpha g_1)\otimes 1, 
\end{equation*}

By the prior relationship this must be of the form: $a (f_1\alpha g_1) \otimes (1\alpha 1)$. However because of Leibniz identity we have $(1\alpha 1) = 0$ and the product $\alpha$ must be trivial. Only by adding another product $\sigma$ we can have non-trivial mathematical structures invariant under system composition.
\subsection{Bipartite products and coproducts}

Let us add another product $\sigma$ and investigate the mathematical consequences. At this point we do not know the nature of the second product $\sigma$. It is helpful to extend the $k$-algebra $(A, \alpha, u)$ to a $k$-two product algebra $(A, \alpha, \sigma, u)$ with $A\otimes A \xrightarrow{\alpha} A$ and $A\otimes A \xrightarrow{\sigma} A$. 

To describe arbitrary system composition we introduce a $k$-coalgebra $C$ as follows:

{\defn Let $C$ be a $k$-space with $\{\alpha, \sigma \}$ as a basis. We define the coproduct $\Delta : C \rightarrow C\otimes C$ and the counit $\epsilon : C \rightarrow k$ as:
\begin{eqnarray*}
\Delta (\alpha) &=& a_{11}~\alpha \otimes \alpha + a_{12}~\alpha \otimes \sigma + a_{21}~\sigma \otimes \alpha + a_{22}~\sigma \otimes \sigma ,\\
\Delta (\sigma) &=& b_{11}~\alpha \otimes \alpha + b_{12}~\alpha \otimes \sigma + b_{21}~\sigma \otimes \alpha + b_{22}~\sigma \otimes \sigma ,\\
\epsilon (\alpha) &=& 0 ,\\
\epsilon (\sigma) &=& 1.
\end{eqnarray*}
} 

We would need to determine the $a$ and $b$ parameters. For quantum mechanics this will become a {\em trigonometric coalgebra} \cite{DascalescuBook}.

\subsection{The fundamental bipartite relationship}

If we consider four elements $f_1 , f_2 , g_1 , g_2$ from the $k$-two product algebra $(A, \alpha, \sigma, u)$, the most general way to construct the products $\alpha$ and $\sigma$ in a bipartite system is as follows:

\begin{eqnarray*}
(f_1 \otimes f_2) \alpha_{12} (g_1 \otimes g_2) = a_{11} (f_1 \alpha g_1) \otimes (f_2 \alpha g_2) + \\ \nonumber
a_{12} (f_1 \alpha g_1) \otimes (f_2 \sigma g_2) + a_{21} (f_1 \sigma g_1) \otimes (f_2 \alpha g_2) + \\ \nonumber
a_{22} (f_1 \sigma g_1) \otimes (f_2 \sigma g_2) , 
\end{eqnarray*}

\begin{eqnarray*}
(f_1 \otimes f_2) \sigma_{12} (g_1 \otimes g_2) = b_{11} (f_1 \alpha g_1) \otimes (f_2 \alpha g_2) + \\ \nonumber
b_{12} (f_1 \alpha g_1) \otimes (f_2 \sigma g_2) + b_{21} (f_1 \sigma g_1) \otimes (f_2 \alpha g_2) + \\ \nonumber
b_{22} (f_1 \sigma g_1) \otimes (f_2 \sigma g_2) . 
\end{eqnarray*}

For convenience we want to normalize the product $\sigma$ such that $1 \sigma f = f \sigma 1 = f$. This can always be done when the product $\sigma$ is distinct from the product $\alpha$. 

By picking appropriate elements and using the fact that $1\alpha 1 = 0$ we can determine some of the $a$ and $b$ parameters. We start by selecting $f_1 = g_1 = 1$. Under this substitution, in $\alpha_{12}$ only terms corresponding to the $a_{21}$ and $a_{22}$ coefficients survive and this demands $a_{21}=1$ and $a_{22} = 0$. Similarly, for $\sigma_{12}$ this demands $b_{21}=0$ and $b_{22}=1$.

Similarly by picking $f_2 = g_2 = 1$ results in $a_{12}=1$ and $b_{12}=0$. In the coalgebra formulation this means:

\begin{equation*}
\Delta(\alpha) = \alpha \otimes \sigma + \sigma \otimes \alpha + a_{11} \alpha \otimes \alpha ,
\end{equation*}
and
\begin{equation*}
\Delta(\sigma) = \sigma \otimes \sigma + b_{11} \alpha \otimes \alpha .
\end{equation*}

We can also prove that $a_{11}=0$. To do this we will use the Leibniz identity on a bipartite system:

\begin{eqnarray*}
(f_1 \otimes f_2) \alpha_{12} [(g_1 \otimes g_2) \alpha_{12} (h_1 \otimes h_2)]=\\ \nonumber
[(f_1 \otimes f_2) \alpha_{12} (g_1 \otimes g_2)] \alpha_{12}(h_1 \otimes h_2) +\\ \nonumber
(g_1 \otimes g_2) \alpha_{12} [(f_1 \otimes f_2)\alpha_{12} (h_1 \otimes h_2)] .
\end{eqnarray*}

Substituting the expression for $\alpha_{12}$ and tracking only the $a_{11}$ terms meaning ignoring any terms involving the $\sigma$ product (because $\alpha$ is a linear product in the $k$-two product algebra) we obtain:

\begin{eqnarray*}
a_{11}^2 [f_1 \alpha (g_1 \alpha h_1)] \otimes [f_2 \alpha (g_2 \alpha h_2)] = \\ \nonumber
a_{11}^2 [(f_1 \alpha g_1 )\alpha h_1)] \otimes [(f_2 \alpha g_2 ) \alpha h_2)] + \\ \nonumber
a_{11}^2 [g_1\alpha (f_1 \alpha h_1)] \otimes [g_2 \alpha (f_2\alpha h_2)] .
\end{eqnarray*}

Applying the Leibniz identity again on the right hand side and canceling terms yields:

\begin{eqnarray*}
a_{11}^2 \{ [(f_1 \alpha g_1) \alpha h_1]\otimes [g_2 \alpha (f_2 \alpha h_2)] + \\ \nonumber
[g_1 \alpha (f_1 \alpha h_1)] \otimes [(f_2 \alpha g_2) \alpha h_2] \} = 0 ,
\end{eqnarray*}

\noindent which is valid for all $f_1 , f_2 , g_1 , g_2$ and hence $a_{11}=0$.

If we rename for convenience $b_{11}$ as $J^2$ (for reasons which will become apparent later) in the end we have the following fundamental relations:
\begin{eqnarray*}
\Delta(\alpha) &=& \alpha \otimes \sigma + \sigma \otimes \alpha ,\\
\Delta(\sigma) &=& \sigma \otimes \sigma + J^2 \alpha \otimes \alpha , 
\end{eqnarray*}

\noindent where $J^2$ can be normalized to be either $-1, 0, +1$. 

Please note the formal similarity with complex number multiplication when $J^2=-1$ which corresponds to quantum mechanics. In this case the coalgebra $C$ is a trigonometric coalgebra \cite{DascalescuBook}. 

$J$ acts as a map from the domain of the product $\alpha$ to the domain of the product $\sigma$. For example (in accordance with Noether's theorem) in quantum mechanics in the Hilbert space representation $J$ is a map from the anti-Hermitean operators which acts as generators of kinematic symmetries to the observables which are Hermitean operators. 

\section{The symmetry properties of the products}

From the composition arguments we have determined three possible solutions for the coproduct which we can call {\em elliptic} for $J^2 = -1$ (quantum mechanics), {\em parabolic} for $J^2 = 0$ (classical mechanics), and {\em hyperbolic} for $J^2 = +1$ for ``hyperbolic quantum mechanics'' \cite{KhrennikovSegre}. 

If the product $\alpha$ is skew-symmetric and the product $\sigma$ is symmetric, the symmetry property is preserved under composition. Just because the product $\alpha$ comes from time evolution it does not necessarily mean it is a skew-symmetric product because it could be a Loday algebra \cite{LodayAlg}. However we can prove that the product $\alpha$ is skew-symmetric: $f\alpha g = -g\alpha f$ by using the Leibniz identity for a bipartite system.

\subsection{The skew-symmetry of the product $\alpha$}

Proving that the product $\alpha$ is skew-symmetric is the essential step which will allow us in the end to reconstruct the algebraic structure of quantum mechanics. In the following we will use the fact that $f\alpha 1 = 1\alpha f = (1 \alpha 1) f = 0 $. Writing down the bipartite Leibniz identity:

\begin{equation*}
f_{12} \alpha_{12} (g_{12} \alpha_{12} h_{12}) = g_{12} \alpha_{12} (f_{12} \alpha_{12} h_{12}) + (f_{12} \alpha_{12} g_{12}) \alpha_{12} h_{12} ,
\end{equation*}

\noindent we observe that on the two right hand side terms $f$'s and $g$'s appear in reverse order and we will want to take advantage of this by carefully choosing the product arguments. Here $f_{12} = f_1 \otimes f_2$ but for typographic convenience in the following we will suppress the tensor symbol and write it as $f_1 f_2$. We select $g_1 = 1 = h_2$ and expand the equation above using the fundamental bipartite relation for product $\alpha$.

Expanding the left hand side we get:

\begin{eqnarray*}
(f_1 f_2)\alpha_{12}[{(g \alpha h)}_1{(g \sigma h)}_2 + {(g \sigma h)}_1{(g \alpha h)}_2] =\\ \nonumber
{(f \alpha (g \alpha h))}_1 {(f \sigma (g \sigma h))}_2 +
{(f \sigma (g \alpha h))}_1 {(f \alpha (g \sigma h))}_2 + \\ 
{(f \alpha (g \sigma h))}_1 {(f \sigma (g \alpha h))}_2 +
{(f \sigma (g \sigma h))}_1 {(f \alpha (g \alpha h))}_2 , 
\end{eqnarray*}

\noindent but this is identically zero because in the first two terms  $g_1 = 1$ and in the last two terms $h_2 = 1$.

The first term on the right hand side expands to:

\begin{eqnarray*}
(g_1 g_2)\alpha_{12}[{(f \alpha h)}_1{(f \sigma h)}_2 + {(f \sigma h)}_1{(f \alpha h)}_2] =\\ \nonumber
{(g \alpha (f \alpha h))}_1 {(g \sigma (f \sigma h))}_2 +
{(g \sigma (f \alpha h))}_1 {(g \alpha (f \sigma h))}_2 + \\
{(g \alpha (f \sigma h))}_1 {(g \sigma (f \alpha h))}_2 +
{(g \sigma (f \sigma h))}_1 {(g \alpha (f \alpha h))}_2 . 
\nonumber
\end{eqnarray*}

In this expression the first and third term vanishes because $g_1 = 1$, and the last term vanish because $h_2 =1$. Because $g_1$ and $h_2$ are units for the product $\sigma$, the overall remaining term is:

\begin{equation*}
{(f \alpha h)}_1{(g \alpha f)}_2 .
\end{equation*}

Finally, the last term on the right hand side expands to:

\begin{eqnarray*}
[{(f \alpha g)}_1{(f \sigma g)}_2 + {(f \sigma g)}_1{(f \alpha g)}_2]\alpha_{12}(h_1 h_2) =\\ \nonumber
{((f \alpha g) \alpha h)}_1 {((f \sigma g) \sigma h)}_2 + 
{((f \alpha g) \sigma h)}_1 {((f \sigma g) \alpha h)}_2 + \\
{((f \sigma g) \alpha h)}_1 {((f \alpha g) \sigma h)}_2 +
{((f \sigma g) \sigma h)}_1 {((f \alpha g) \alpha h)}_2 .  
\nonumber
\end{eqnarray*}

In this expression the first two terms vanish because $g_1 = 1$, and the last term vanishes because $h_2 =1$. Because $g_1$ and $h_2$ are units for the product $\sigma$, the overall remaining term is:

\begin{equation*}
{(f \alpha h)}_1{(f \alpha g)}_2 .
\end{equation*}

Putting it all together yields:

\begin{equation*}
0 = {(f \alpha h)}_1[{(f \alpha g)}_2 + {(g \alpha f)}_2] ,
\end{equation*}
 
\noindent which is valid for any arbitrary ${(f \alpha h)}_1$ terms. Hence:

\begin{equation*}
f \alpha g = -g \alpha f ,
\end{equation*}

\noindent and the skew-symmetry of the product $\alpha$ is proved.

\subsection{The symmetry of the product $\sigma$}

To prove that $f\sigma g = g \sigma f$ we will use the fundamental bipartite relationship for $\alpha_{12}$ and the just proved skew-symmetry of $\alpha$.

The bipartite expression for the product $\alpha$ reads:

\begin{equation*}
(f_1 f_2) \alpha_{12} (g_1 g_2) = {(f\alpha g)}_1 {(f\sigma g)}_2 + {(f\sigma g)}_1 {(f\alpha g)}_2 .
\end{equation*}

This is also equal with:

\begin{equation*}
-(g_1 g_2) \alpha_{12} (f_1 f_2) = -{(g\alpha f)}_1 {(g\sigma f)}_2 - {(g\sigma f)}_1 {(g\alpha f)}_2 ,
\end{equation*} 

\noindent and

\begin{equation*}
-(g_1 g_2) \alpha_{12} (f_1 f_2) = {(f\alpha g)}_1 {(g\sigma f)}_2 + {(g\sigma f)}_1 {(f\alpha g)}_2 .
\end{equation*}

We therefore have:

\begin{equation*}
{(f\alpha g)}_1 {[(f\sigma g) - (g\sigma f)]}_2 + {[(f\sigma g) - (g\sigma f)]}_1{(f\alpha g)}_2 = 0 .
\end{equation*}

Suppose now that we pick the functions $f$ and $g$ such that ${(f\alpha g)}_1 \neq 0$ and ${(f\alpha g)}_2 \neq 0$. We then have:

\begin{equation*}
1 \otimes \frac{{[(f\sigma g) - (g\sigma f)]}_2}{{(f\alpha g)}_2} + \frac{{[(f\sigma g) - (g\sigma f)]}_1}{{(f\alpha g)}_1} \otimes 1 = 0 .
\end{equation*}

The only way system $1$ value can be equal with system $2$ value is if both expressions are equal with a constant $c$:

\begin{equation*}
1 \otimes c + c \otimes 1 = 0 .
\end{equation*}

However by using the identity property for the tensor product this means that $c+c=0$ and hence $c=0$. In turn this demands the symmetry of the product $\sigma$: $(f\sigma g) = (g\sigma f)$.

\section{The complete algebraic properties of the products}

Now we can establish that the product $\alpha$ forms a Lie algebra satisfying the Jacobi identity, and that the product $\sigma$ is a Jordan product. Together they obey a compatibility relationship allowing the definition of an associative product.

\subsection{The Lie algebra}

The product $\alpha$ is linear in the second term because $(f \alpha \cdot)$ is a derivation, is skew-symmetric, and respects the Leibniz identity:

\begin{equation*}
f \alpha (g \alpha h) = (f \alpha g) \alpha h + g \alpha (f \alpha h) .
\end{equation*}

By the skew-symmetry property we get:

\begin{equation*}
f \alpha (g \alpha h) = -h \alpha (f \alpha g) - g \alpha (h \alpha f) ,
\end{equation*}

\noindent which is the Jacobi identity. Hence $\alpha$ is a Lie algebra. 

\subsection{The compatibility relationship and the Jordan identity}

The two products $\alpha$ and $\sigma$ are not independent and they obey a compatibility relation:

\begin{equation*}
{[ f, g, h] }_{\sigma} + J^2 {[f, g, h]}_{\alpha} = 0,
\end{equation*}

\noindent where $[f, g, h]_{\circ}$ is {\em the associator} of the product $\circ$: 
\begin{equation*}
[f, g, h]_{\circ} = (f\circ g) \circ h - f \circ (g \circ h).
\end{equation*}

In Hilbert space formulation this arises out of the C*-algebra but this relationship can again be obtained from composition considerations. Using the assumptions of the symmetry of the product $\sigma$, the skew-symmetry of the product $\alpha$, the Jacobi identities, and the fundamental bipartite relations the proof of the compatibility relation was first obtained by Grgin and Petersen \cite{GP}. Because the proof is not new, we will only sketch it here for completeness sake. Grgin and Petersen start from the bipartite Jacobi identity:

\begin{equation*}
\sum_{\rm cycl} (f_1 f_2)\alpha_{12} ((g_1 g_2)\alpha_{12}(h_1 h_2)) = 0 .
\end{equation*}

After expansion and usage of the Leibniz identity, it becomes:

\begin{equation*}
\sum_{\rm cycl} {(f \sigma (g \sigma h))}_1 {(f \alpha (g \alpha h))}_2 + {(f \alpha (g \alpha h))}_1 {(f \sigma (g \sigma h))}_2 = 0 .
\end{equation*}

Adding it to a copy of itself but with $g_1$ and $h_1$ interchanged results in:
\begin{eqnarray*}
\{[f, g, h]_{\sigma} + [f, h, g]_{\sigma}\}_1 \{f \alpha (g \alpha h)\}_2 = \\
\{(g \alpha (h \alpha f)) + (h \alpha (g \alpha f))\}_1 \{[h, f, g]_{\sigma}\}_2 ,
\end{eqnarray*}

This implies a relation of proportionality:

\begin{equation*}
(f \alpha (g \alpha h)) = \lambda [h, f, g]_\sigma.
\end{equation*}

Using the Jacobi identity on the left hand side, it yields the compatibility relation:
\begin{equation*}
{[ f, g, h] }_{\sigma} + \lambda {[f, g, h]}_{\alpha} = 0.
\end{equation*}

The remaining part of the proof is establishing the relation between $\lambda$ and $J^2$ which occurs in the bipartite expansion of the product $\sigma_{12}$. To this aim Grgin and Petersen use the bipartite Leibniz identity to expand:

\begin{equation*}
(f_1 f_2)\alpha_{12} ((g_1 g_2)\sigma_{12}(h_1 h_2)) ,
\end{equation*}
\noindent and working along similar lines as above they derive a proportionality property which this time involves $J^2$. In the end the compatibility identity is obtained.

The Jordan identity is now a straightforward consequence of the compatibility identity when $f,g,h$ are chosen to be $f, g, f\sigma f$ respectively. With this choice the $\alpha$ associator is zero:

\begin{eqnarray*}
[f, g, f^2]_\alpha = (f \alpha g) \alpha (f^2) - f \alpha (g \alpha f^2) = \\
(f \alpha g) \alpha (f^2) - (f \alpha g) \alpha (f^2) - g \alpha (f \alpha f^2) = 0 .
\end{eqnarray*}

The last term $g \alpha (f \alpha f^2)$ is zero because $f \alpha f^2 = f \alpha (f \sigma f) = (f\alpha f) \sigma f + f \sigma (f \alpha f) = 0$. Hence from the compatibility relationship we have: $[f, g, f^2]_\sigma = 0$ which is another formulation of the Jordan identity (power associativity).

\subsection{The associative product}

To arrive at states and the usual formalism of quantum mechanics one needs an additional ingredient, an associative multiplication $\beta = \sigma \pm J \alpha$ 

Associativity follows from the associator property of the composability two-product algebra. However each product appears twice and the proof is not obvious.

Let us compute the associator ${[f, g, h]}_{\beta} = (f \beta g) \beta h - f \beta ( g \beta h)$ using the definition of $\beta$:

\begin{eqnarray*}
{[f, g, h]}_{\beta} = (f \sigma g \pm J f \alpha g) \beta h  - f \beta (g \sigma h \pm J g \alpha h)\\ \nonumber
=  (f \sigma g) \sigma h \pm J (f \sigma g ) \alpha h \pm J (f \alpha g) \sigma h + J^2 (f \alpha g) \alpha h \\ \nonumber
- f \sigma (g \sigma h) \mp J f \sigma (g \alpha h) \mp J f \alpha (g \sigma h) - J^2 f \alpha (g \alpha h) \\ \nonumber
= {[f, g, h]}_{\sigma} + J^2 {[f, g, h]}_{\alpha} \\ \nonumber
\pm J \{(f \sigma g) \alpha h + (f\alpha g) \sigma h - f \sigma (g \alpha h) - f\alpha (g \sigma h)\} = 0 .
\end{eqnarray*}

In the last line the terms cancel after using the Leibniz rule for $f\alpha (g\sigma h)$ and $(f \sigma g) \alpha h$.

Because $\beta$ is an associative product and $\sigma$ corresponds to its real part, the Jordan algebra of observables $\sigma$ cannot be special. 

\subsection{The algebraic structure of quantum and classical mechanics}

Now we can collect all the result above and introduce the composability two-product algebra which forms the algebraic structure of quantum and classical mechanics. To make it identical with the usual products we redefine the product $\alpha$ such that the map $J$ becomes $J \hbar/2$.

{\defn A composability two-product algebra is a real vector space $\mathfrak{A}_{\RR}$ equipped with two bilinear maps $\sigma$ and $\alpha$ such that the following conditions apply:
\begin{eqnarray*}
\alpha {\rm ~is~a~Lie~algebra} ,\\
\sigma {\rm ~is~a~Jordan~algebra} ,\\
\alpha {\rm ~is~a~derivation~for~}\sigma {\rm ~and~} \alpha ,\\
{[ A, B, C] }_{\sigma} + \frac{J^2 \hbar^2}{4} {[A, B, C]}_{\alpha} = 0 ,
\end{eqnarray*}
where $J \rightarrow (-J)$ is an involution, $1\alpha A = A\alpha 1 = 0$, $1\sigma A = A\sigma 1 = A$, and $J^2 = -1,0,+1$. 
} 

Quantum mechanics corresponds to $J^2 = -1$ (elliptic composability), classical mechanics corresponds to $J^2 = 0$ (parabolic composability), and the unphysical hyperbolic quantum mechanics corresponds to $J^2 = +1$ (hyperbolic composability). Hyperbolic quantum mechanics is unphysical becase the Stone-von Neumann theorem does not hold and one cannot generate physical predictions independent of the position or momentum representation \cite{KhrennikovSegre}.

For classical mechanics, when $J^2 = 0$, the product $\sigma$ become associative, which is stronger then power associative, and in fact $\sigma$ is simply the regular function multiplication. Considerations of norm do not enter in the algebraic structure and the composability two-product algebra is simply a Jordan-Lie algebra without Banach norm axioms.

\section{Representations of the composability two-product algebra}

It is informative to present now the concrete representations of the composability two-product algebra and the reader can easily check that the satisfy all the algebraic properties.

Starting with classical mechanics, in phase space the product $\sigma$ is the regular function multiplication, while the product $\alpha$ is the Poisson bracket:

\begin{equation*}
f \alpha g = \{ f , g \} = f \overleftrightarrow{\nabla} g = \sum_{i = 1}^{n}  \frac{\partial f}{\partial q^i} \frac{\partial g}{\partial p_i} - \frac{\partial f}{\partial p_i} \frac{\partial g}{\partial q^i} .
\end{equation*}

Quantum mechanics has a Hilbert space representation, and a phase space representation. The algebraic products of the Hilbert space representation are as follows:

\begin{eqnarray*}
A\alpha B &=& \frac{J}{\hbar}(AB - BA), \\ \nonumber
A\sigma B &=& \frac{1}{2} (AB + BA) ,
\end{eqnarray*}
\noindent which are the usual commutator and the Jordan product. Here $J$ is the complex number imaginary unit: $i = \sqrt{-1}$. The associative product $\beta = \sigma - \frac{J\hbar}{2} \alpha$ is the usual operator multiplication.

In (flat) phase space formulation the products $\alpha$ and $\sigma$ are the Moyal and the cosine brackets \cite{MoyalBracket}:

\begin{eqnarray*}
\alpha &=& \frac{2}{\hbar}\sin ( \frac{\hbar}{2}\overleftrightarrow{\nabla}) ,\\ \nonumber
\sigma &=& \cos (\frac{\hbar}{2}\overleftrightarrow{\nabla}) ,
\end{eqnarray*}

\noindent where the operator $\overleftrightarrow{\nabla}$ is defined as follows:

\begin{equation*}
\overleftrightarrow{\nabla} = \sum_{i=1}^{N} [ \overleftarrow{\frac{\partial}{\partial x^i}}\overrightarrow{\frac{\partial}{\partial p_i}} - \overleftarrow{\frac{\partial}{\partial p_i}}\overrightarrow{\frac{\partial}{\partial x^i}}] .
\end{equation*}

The associative product $\beta = \sigma + \frac{J\hbar}{2} \alpha$ is the star product:

\begin{equation*}
f \star g = f \sigma g + \frac{J\hbar}{2} f \alpha g = f e^{\frac{J\hbar}{2} \overleftrightarrow{\nabla}} g .
\end{equation*}

\section{Conclusion}

From very general composition arguments we were able to obtain three composition classes classified by the parameter $J^2$. There is no cross talk between the classes \cite{SahooPlanck} and in particular there can be no consistent theory of Nature which combine classical and quantum mechanics. Nature can only be in one of the composition classes, and the way to determine it is by experimental evidence \cite{AspectExperiment}. 

In the finite dimensional case, the composability two-product algebra is enough to fully reconstruct quantum mechanics by an appeal to Artin-Wedderburn theorem \cite{AKapustin}. The infinite dimensional case is harder and one has to add positivity and norm considerations, like in the C*-algebra condition:

\begin{equation*}
||x^* x|| = {||x||}^2 .
\end{equation*}  

An additional consideration is that of the number system. It is well known that quantum mechanics can be expressed over real numbers, complex, or quaternionic numbers \cite{AdlerQuaternions} and this correspond to representations of transition probabilities \cite{LandsmanBook}. However, if probabilities are generalized to four-vector relativistic current probabilities subject to a continuity condition, then the number system generalizes to $SL(2, \CC)$ and the C*-algebra becomes a C*-Hilbert module \cite{GrginBook}. The resulting theory is a gauge theory with the gauge group $U(1)\times SU(2)$. This shows that the full classification of the representations of the composability two-product algebra is a much harder problem. Quaternionic and real quantum mechanics do not admit tensor composition. If we demand the existence of the tensor product as well as transition probabilities then we single out complex quantum mechanics.


\bibliography{aipAlgebraicStructure}

\end{document}